\documentclass{jfm}
\usepackage{graphicx}
\usepackage{natbib}
\usepackage{graphicx,float,calc,ifthen,amsmath}
\ifCUPmtlplainloaded \else
  \checkfont{eurm10}
  \iffontfound
    \IfFileExists{upmath.sty}
      {\typeout{^^JFound AMS Euler Roman fonts on the system,
                   using the 'upmath' package.^^J}%
       \usepackage{upmath}}
      {\typeout{^^JFound AMS Euler Roman fonts on the system, but you
                   dont seem to have the}%
       \typeout{'upmath' package installed. JFM.cls can take advantage
                 of these fonts,^^Jif you use 'upmath' package.^^J}%
       \providecommand\upi{\pi}%
      }
  \else
    \providecommand\upi{\pi}%
  \fi
\fi

\ifCUPmtlplainloaded \else
  \checkfont{msam10}
  \iffontfound
    \IfFileExists{amssymb.sty}
      {\typeout{^^JFound AMS Symbol fonts on the system, using the
                'amssymb' package.^^J}%
       \usepackage{amssymb}%

      }{}
  \fi
\fi

\ifCUPmtlplainloaded \else
  \IfFileExists{amsbsy.sty}
    {\typeout{^^JFound the 'amsbsy' package on the system, using it.^^J}%
     \usepackage{amsbsy}}
    {\providecommand\boldsymbol[1]{\mbox{\boldmath $##1$}}}
\fi
\usepackage{hyperref}      
\hypersetup{
    colorlinks=true,       
    linkcolor=blue,          
    citecolor=blue,        
    filecolor=blue,      
    urlcolor=blue          
}

\newcommand\Pen{\mbox{\textit{Pe}}}  

\newsavebox{\astrutbox}
\sbox{\astrutbox}{\rule[-5pt]{0pt}{20pt}}

\usepackage[usenames,dvipsnames]{color}

\title[Revised non-normality in combustion-acoustic interaction in diffusion flames]{Non-normality in combustion-acoustic interaction in diffusion flames: a critical revision\thanks{This is a pre-print version. Published in J.\ Fluid\ Mech., vol. 733, pp. 681--683, 2013. Cambridge University Press$^{\copyright}$. DOI: 10.1017/jfm.2013.468}}

\author[L. Magri, K. Balasubramanian, R.I. Sujith and M.P. Juniper]
{Luca Magri$^1$\thanks{Email address for correspondence: \href{mailto:lm547@cam.ac.uk}{lm547@cam.ac.uk}}
, 
K. Balasubramanian$^2$ 
,
R. I. Sujith$^3$ 
and 
M. P. Juniper$^1$}

\affiliation{$^1$Department of Engineering, University of Cambridge,\\
Trumpington Street, Cambridge, CB2 1PZ, UK\\[\affilskip]
$^2$C.N. Yang Institute for Theoretical Physics, 
	                           State University of New York,\\
	                           Stony Brook, NY 11794-3840, 
	                           USA\\[\affilskip]
$^3$Department of Aerospace Engineering, Indian Institute of Technology Madras,\\Chennai 600036, Tamil Nadu, India}

\pubyear{xxxx}
\volume{xxx}
\pagerange{xxx--xxx}
\date{?? and in revised form ??}
\begin{document}
\maketitle
%
%
%
%
\begin{abstract}
Perturbations in a non-normal system can grow transiently
even if the system is linearly stable.
If this transient growth is sufficiently large,
it can trigger self-sustained oscillations from small initial disturbances.
This has important practical consequences for combustion-acoustic oscillations,
which are a continual problem in rocket and aircraft engines.  
Balasubramanian and Sujith ({\it Journal of Fluid Mechanics} 2008, 594, 29--57) modelled an infinite-rate chemistry diffusion flame in an acoustic duct and found that the transient growth in this system can amplify the initial energy by a factor, $G_{max}$, of order $10^5$ to $10^7$. 
However,
recent investigations by L. Magri \& M. P. Juniper 
have brought to light certain errors in that paper. 
When the errors are corrected, $G_{max}$ is found to be of order 1 to 10,
revealling that non-normality is not as influential as it was thought to be. 
\end{abstract}
%
\section{Results and discussion}
In this note,
we use the same model, discretization, and non-dimensionalization as \citet{BALASUB-08b,BALASUB-13}, labelled B\&S for brevity,
but include the corrected equations, which are listed below. It is implied that the following errata refer to B\&S.
\vspace{10mm}
%
\begin{enumerate}
\item The analytical  steady solution, $Z_{st}$ (appendix B, p. 54), obtained by separation of variables, is:
\begin{equation}\label{zstind}
Z_{st}=X_i\left(1-\alpha\right)
-Y_i\alpha 
-\frac{2}{\upi}\left(X_i+Y_i\right)\sum_{n=1}^{+\infty}\frac{\sin(n\upi\alpha)}{n\left(1+b_n\right)}\cos(n\upi y_c)\left(e^{a_{n1}x_c}+b_n e^{a_{n2}x_c}\right),
\end{equation}
where 
\begin{align}
&a_{n1}\equiv\frac{\Pen}{2}-\sqrt{\frac{\Pen^2}{4}+n^2\upi^2},\;\;\;\; a_{n2}\equiv\frac{\Pen}{2}+\sqrt{\frac{\Pen^2}{4}+n^2\upi^2},\label{zstind2}\\
&b_n\equiv-\frac{a_{n1}}{a_{n2}}e^{\left(-2L_c\sqrt{\frac{\Pen^2}{4}+n^2\upi^2}\right)}.\label{zstind3}
\end{align}
The non-dimensional coordinates of the combustion domain are $x_c$, $y_c$.
%
%
%
%
%
\item  The expressions for $C^{(n)}_m$ and $W_{mk}$ (eq.~(7), p. 36) are $2/L_c$ times the original terms due to Galerkin projection; and $W_{mk}=1/L_c$ when $k= m$.
\item The variable $Y_1$ in the right-hand side of the terms $R_{nm}$ and $J_{nm}$ (eq.~(13), p. 37) is $Y_i$. 
\item $\dot{Q}_{av}$ (eqs.~(18),(19), p. 39) is to be divided by 2 due to non-dimensionalization over the cross-sectional area. 
\item The multiplying factor ahead matrix $\mathsfbi{M_1}$ (appendix B, p. 54) is $\frac{1}{(T_i+T_{ad})/2}$.
\item The expression of the matrix $\mathsfbi{B}_{NN}$ (appendix B, p. 54) is $\mathsfbi{B}_{NN}=-\mathsfbi{D} + \mathsfbi{A}_1- \mathsfbi{A}_2 + \mathsfbi{A}_3- \mathsfbi{A}_4+ \mathsfbi{A}_5$, where
\begin{equation}
\mathsfbi{A}_5 = \frac{1}{(T_{i} + T_{ad})/2}[0\;0\;0\;\ldots\;0\;\sin(\upi x_f)\;\sin( 2 \upi x_f)\;\ldots\;\sin(K \upi x_f)]^T [J_{00}\;\ldots\;J_{0M}\;0\;\ldots\;0].
\end{equation}
$K$ is the number of Galerkin modes for acoustic discretization.
\item  The damping terms in the matrix $\mathsfbi{S}$ (appendix B, p. 55) are
$+2\upi\xi_1, +4\upi\xi_2, \ldots, +2K\upi\xi_K$. 
\item The numerator of the matrix $\mathsfbi{A}_4$ (appendix B, p. 55) is 1 due to non-dimensionalization over the cross-sectional area of the duct. 
\end{enumerate}
\vspace{10mm}
Computations are performed by using $50 \times 50$ Galerkin modes in the flame domain, and 6 modes in the acoustic domain. 
When the number of Galerkin modes is increased to $70\times 70$ in the flame and 12 in the acoustics, 
the eigenvalues and singular values change by less than $15\%$.
The fixed parameters are the fuel mass ratio, $Y_i=3.2$; the oxidizer mass ratio, $X_i=3.2/7$; and the average temperature, $T_{av}=1/0.685$.
We set the damping coefficients to $c_1=0.013$ and $c_2=0.08$ in order to have marginally stable systems.
The nonlinear behaviour of this thermo-acoustic system is not considered because it has been fully characterized by \citet{ILLING-13}. 

\begin{figure}
\begin{center}
\includegraphics[width=0.9\textwidth, draft = false]{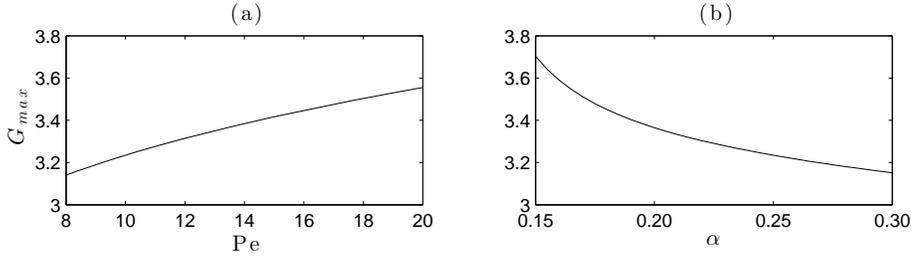}
\caption{The growth factor, $G_{max}$, as a function of (a) the P\'eclet number, $\Pen$, and (b) the fuel slot half width, $\alpha$. Frame (a) has $\alpha = 0.25$, frame (b) has $\Pen=10$.} 
\label{fig:1}
\end{center}
\end{figure}

Figures~\ref{fig:1}a,b show the growth factor, $G_{max}$\footnote{We use the same norm as B\&S, even though Chu's norm would be a more appropriate measure of the energy \citep{CHU-65}}, as a function of the P\'eclet number, $Pe$, and the non-dimensional half width of the fuel slot, $\alpha$, respectively.
In both cases, $1<G_{max}\lesssim 10$.
Furthermore, marginally stable but highly non-normal fluid-dynamic systems exhibit pseudospectra that protrude significantly into the unstable half-plane \citep{TREF-05}. 
In this thermo-acoustic system, however, the pseudospectra around the most unstable eigenvalues are nearly concentric circles whose values decrease rapidly as the distance from the eigenvalue increases. 
This is a further demonstration that the system is only weakly non-normal. 
It is worth noting, however, that \cite{JUNIPER-11} showed that even a small amount of non-normality can make a system somewhat more susceptible to triggering. \\

L.M. and M.P.J. would like to thank Iain Waugh for valuable discussions and Alessandro Orchini for scrutinizing some parts of the  code. L.M. is supported by the European Research Council through Project ALORS 2590620. 
\begin{figure}
\begin{center}
\includegraphics[width=0.5\textwidth, draft = false]{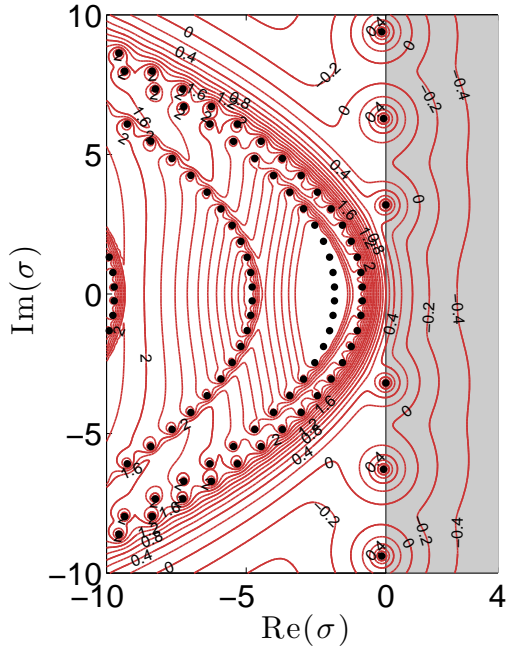}
\caption{Logarithm of the pseudospectra, $\log_{10}(\epsilon)$.
The parameters are the same as figure~\ref{fig:1}, with $\alpha=0.25$ and $\Pen=10$. The dominant eigenvalue is $\sigma =-0.003 \pm 3.193\mathrm{i}$.
}. 
\label{fig:2}
\end{center}
\end{figure}
%
%
%
%
%
%

%
%
%
%
\end{document}